\documentclass[aip,reprint,pop,amsmath,amssymb]{revtex4-1}

\usepackage{graphicx}
\usepackage{dcolumn}
\usepackage{bm}

\begin{document}

\title{Nonlinear stability of the ideal magnetohydrodynamic interchange mode at marginal conditions in a transverse magnetic field}

\author{Jupiter Bagaipo}
\email{jbagaipo@umd.edu}
\author{P. N. Guzdar}
\author{A. B. Hassam}
\address{Institute for Research in Electronics and Applied Physics, University of Maryland, College Park, MD 20742-3511}

\date{18 July 2011}

\begin{abstract}
The stability of the ideal magnetohydrodynamic (MHD) interchange mode at marginal conditions is studied. A sufficiently strong constant magnetic field component transverse to the direction of mode symmetry provides the marginality conditions. A systematic perturbation analysis in the smallness parameter, $|b_2/B_c|^{1/2}$, is carried out, where $B_c$ is the critical transverse magnetic field for the zero-frequency ideal mode, and $b_2$ is the deviation from $B_c$. The calculation is carried out to third order including nonlinear terms. It is shown that the system is nonlinearly unstable in the short wavelength limit, i.e., a large enough perturbation results in instability even if $b_2/B_c>0$ (linearly stable). The normalized amplitude for instability is shown to scale as $|b_2/B_c|^{1/2}$. A nonlinear, compressible, MHD simulation is done to check the analytic result. Good agreement is found, including the critical amplitude scaling.
\end{abstract}

\maketitle

\section{\label{sec:intro}Introduction}
It is well known that the magnetohydrodynamic (MHD) magnetized plasma interchange instability can be stabilized by a transverse magnetic field. For a given wavenumber, allowing a magnetic field component in the direction of the wavenumber introduces Alfv\'{e}nic stabilizing tension such that beyond a critical transverse field (transverse to the direction of mode symmetry) that wavenumber is linearly stable.\cite{Freidberg-book} The nonlinear evolution of the magnetized plasma interchange instability is less well understood. In particular, the state of the system for when the transverse B-field is marginally subcritical (or, equivalently, the plasma beta is slightly above critical) is an important question for magnetized fusion energy applications: does the mode saturate at low amplitude and how does the marginal convection and resulting transport scale with deviation from marginality? The question is an important consideration for stellarators, for example, since these fusion devices are engineered for very high precision magnetic fields and one of the precision constraints arises from ideal MHD linear stability results.\cite{Kulsrud} If the nonlinear consequence of a slightly subcritical B-field were better understood, it may be possible to optimize over the MHD design constraints. It was also recently shown that the linear growth rate of ideal interchanges in a reversed-field pinch for a slightly subcritical B-field is weaker than expected and may be overcome by nonlinear effects.\cite{Gupta}

The interchange mode is a pressure-driven mode that is characterized by the interchange of magnetic flux tubes so that the overall free energy of the system is lowered.\cite{Longmire} The instability occurs when the equilibrium has a density gradient unfavorable to the direction of a ``gravitational'' force. In systems with curvature, this force comes from a centrifugal force generated by thermal motion in field curvature. The mode can be stabilized by introducing a strong enough field, transverse to the flutes, that prevents the flux tubes from being able to freely interchange. The strength of the stabilizing field can be determined using linear theory and will depend on the steepness of the density gradient and the magnitude of the gravitational force.

There have been a few studies done on nonlinear growth of interchange instabilities at marginal stability in tokamaks.\cite{Rutherford,Waelbroeck,Beklemishev,Cowley,Zhu2,Zhu1} Although a Lagrangian approach has been attempted,\cite{Pfirsch} the general approach is to expand the equation of motions of the unstable mode about marginal stability and thus the nonlinear terms in the system can be evaluated.\cite{Waelbroeck,Beklemishev} We can determine the overall stability of the system by comparing the behaviour of the nonlinear effects to the linear driving term. In Ref.~\onlinecite{Waelbroeck} the author found that, for the profiles investigated, the nonlinear effects were stabilizing. Similarly, in Ref.~\onlinecite{Beklemishev} the author showed nonlinear saturation at marginal stability. Both authors considered a system with a sheared magnetic field. In studying the line-tied $g$ mode, the authors in Ref.~\onlinecite{Cowley} showed that near the marginally stable point the system was nonlinearly unstable. However, Refs.~\onlinecite{Zhu2,Zhu1} showed that the nonlinear growth transitions through an initial regime where the nonlinear growth dominates the linear response, as shown in Ref.~\onlinecite{Cowley}, but a secondary regime takes over when the amplitude is sufficiently large and so the mode amplitude remains bounded.

We simplify our system to a slab geometry where we use an effective gravitational field, $\mathbf{g}$, to model centrifugal force due to field line curvature\cite{Longmire} and we assume a constant transverse field. This reduces the complexity of the system so that the focus of the analysis can be on how nonlinear terms get introduced into the equations of motion. The idealized system is described in Sec.~\ref{sec:theory} along with the derivation of nonlinear time evolution equation. The goal is to have a simpler methodology in a simple system that can be generalized into more complicated systems, e.g. sheared field\cite{Suydam}, ballooning\cite{Drake,Connor}, etc. In Sec.~\ref{sec:num} we verify our result using a dissipative numerical simulation. The results are summarized in Sec.~\ref{sec:summary}.

\section{\label{sec:theory}Theory}
Consider a slab system with constant gravity $\mathbf{g}=-g\mathbf{\hat{x}}$ and very strong magnetic field in the $z$-direction such that $B_\perp \ll B_z$ and $V_{Az}$ is much larger than $u$, the typical flow in the system. This system is incompressible and can be described by the two-dimensional MHD reduced equations\cite{Strauss} given by,
\begin{equation}\label{eq:den}
\partial_t \rho + \mathbf{u}\cdot\mathbf{\nabla}_\perp \rho = 0,\end{equation}
\begin{equation}\label{eq:phi}
\mathbf{\hat{z}}\cdot\mathbf{\nabla}_\perp\times(\rho\frac{d}{dt}\mathbf{u}) 																								=\mathbf{B}\cdot\mathbf{\nabla}_\perp\nabla_\perp^2\psi+g\partial_y \rho,\end{equation}
\begin{equation}\label{eq:psi}
\partial_t \psi - \mathbf{B}\cdot\mathbf{\nabla}_\perp\phi = 0,\end{equation}
where $\mathbf{u}=\mathbf{\hat{z}}\times\mathbf{\nabla}_\perp\phi$ and $\mathbf{B_\perp}=\mathbf{\hat{z}}\times\mathbf{\nabla}_\perp\psi$ and we have defined, in the usual way,
\begin{equation*}
\frac{d}{dt}\equiv\frac{\partial}{\partial t}+\mathbf{u}\cdot\mathbf{\nabla}_\perp.
\end{equation*}
Variations in $z$ are suppressed since the fastest interchange has $\partial/\partial z=0$. The nonlinear system of equations (\ref{eq:den})-(\ref{eq:psi}) can be solved for the variables $\rho$, the density, and $\phi$ and $\psi$, the flow and magnetic streamfunctions, respectively.

We consider a static equilibrium with a density gradient that's unstable to interchange and a constant, transverse magnetic field. More explicitly, we have
\begin{equation}\label{equil}
\rho_0'(x) > 0, \quad \mathbf{B}=B_0\mathbf{\hat{y}}, \quad \phi_0=0,
\end{equation}
where henceforth the primes denote differentiation with respect to $x$. We also add the assumption that $\rho_0'\rightarrow 0$ at the boundaries and has even parity.

Small perturbations about this equilibrium yield the WKB dispersion relation 
\begin{equation}\label{eq:disp}
\omega^2 = k^2V_{Ay}^2-\gamma_g^2
\end{equation}
where $\gamma_g=|g\rho'/\rho|^{1/2}$ is the Rayleigh-Taylor growth rate and $k$ is the wavenumber in the $y$ direction. The Rayleigh-Taylor growth rate represents the effective ``gravitational'' acceleration and is the driving force in an interchange instability. With a transverse magnetic field, field line bending results in Alfv\'{e}nic restoring forces with frequency $k V_{Ay}$. In this paper, we consider the dynamics of the magnetized interchange mode when the magnetic field strength is strong enough to just stabilize interchanges, i.e., the system is near marginal stability. In particular, for a given $k$, suppose $\omega^2>0$ everywhere in $x$ except for a single small region where it is very close to zero, positive or negative. In that case, weakly growing perturbations are possible in the vicinity of where $k^2V_{Ay}^2-\gamma_g^2$ is close to zero. The time rate of change of the perturbations will be very small compared to the local $\gamma_g$. Thus, we order
\begin{equation}
\partial_t/\gamma_g\sim\epsilon\ll 1.
\end{equation}
This implies that any deviations in $B_0$ away from criticality must be small. In particular, if
\begin{equation}\label{eq:B_0}
B_0=B_c+b_2
\end{equation}
then, according to (\ref{eq:disp}), $b_2/B_c$ must be of $\mathcal{O}(\epsilon^2)$.

We allow small perturbations about this marginal point such that the amplitude of the magnetic perturbation, $A$, while small, is large enough that the nonlinear magnetic tension forces can influence the growth time. This results in the optimal ordering
\begin{equation}
A/\psi_0\sim\epsilon.
\end{equation}
We represent the perturbation by expanding $\psi$ and $\phi$ in a series
\begin{equation}\label{eq:expsi}
\psi=\psi_0+\psi_1+\psi_2+\psi_3+\cdots
\end{equation}
\begin{equation}\label{eq:exphi}
\phi=\phi_1+\phi_2+\phi_3+\cdots
\end{equation}
where the order in $\epsilon$ is denoted by the subscript. The continuity equation (\ref{eq:den}) can be satisfied by letting $\rho=\rho(\psi)$ and using (\ref{eq:psi}). With this change of variable we can expand $\rho$ in terms of $\delta\psi=\psi-\psi_0$ to get
\begin{equation}\label{eq:exden}
\rho(\psi)=\rho_0+\frac{\rho_0'}{B_0}\delta\psi+\frac{1}{2}\frac{\rho_0''}{B_0^2}\delta\psi^2+\frac{1}{6}\frac{\rho_0'''}{B_0^3}\delta\psi^3+\cdots.
\end{equation}
Substituting the expansions (\ref{eq:expsi})-(\ref{eq:exden}) into (\ref{eq:phi}) and (\ref{eq:psi}) we can solve for the nonlinear evolution of the perturbations order by order.

\subsection{\label{sec:first}First order equations}
Matching terms to lowest, non-vanishing order, we obtain from (\ref{eq:phi}) and (\ref{eq:psi}) the equations
\begin{equation}\label{eq:phi1}
0=B_c\partial_y\nabla_\perp^2\psi_1+g\partial_y\rho_1,
\end{equation}
\begin{equation}\label{eq:psi1}
-B_c\partial_y\phi_1=0,
\end{equation}
where (\ref{eq:exden}) gives
\begin{equation}\label{eq:den1}
\rho_1=\rho_0'\psi_1/B_c.
\end{equation}
Substituting $\rho_1$ into (\ref{eq:phi1}) the equation becomes
\begin{equation}\label{eq:phi12}
\mathcal{L}(\psi_1)=0
\end{equation}
where we have defined the operator
\[
\mathcal{L}(f)\equiv(\nabla_\perp^2+\frac{g}{B_c^2}\rho_0')\partial_yf.
\]
Writing $\psi_1$ as
\begin{equation}\label{eq:psi1sol}
\psi_1(x,y,t)=A(t)\zeta_1(x)\cos(ky),
\end{equation}
we obtain the eigenvalue equation
\begin{equation}\label{eq:eig}
\zeta_1''(x)-k^2\zeta_1(x)+\frac{g}{B_c^2}\rho_0'\zeta_1(x)=0
\end{equation}
that can be solved to get an eigenvalue for $B_c$. The boundary condition for $\rho_0$ implies that $\zeta_1$ decays exponentially close to the boundary.

In writing (\ref{eq:psi1sol}) we assumed a cosine perturbation in the density which implies $\psi_1\sim\cos(ky)$ from (\ref{eq:exden}). If we also assume that this initial perturbation results in a pure mode for the lowest order flow then
\begin{equation}\label{eq:phi1sol}
\phi_1(x,y,t)=0
\end{equation}
is the solution to (\ref{eq:psi1}).

To the lowest order we have found that given the mode of the density perturbation, $k$, and the equilibrium density gradient profile, $\rho_0'(x)$, then the marginally stable field strength $B_c$ can be solved for using (\ref{eq:eig}). This result is consistent with the prediction from linear theory for the existence of the marginally stable value.

\subsection{\label{sec:sec}Second order equations}
In order to solve for the time evolution of $\psi_1$, it is necessary to proceed to higher order in the expansion. We now match $\mathcal{O}(\epsilon^2)$ terms in equations (\ref{eq:phi}) and (\ref{eq:psi}) to get
\begin{equation}\label{eq:phi2}
B_c\partial_y\nabla_\perp^2\psi_2+g\partial_y\rho_2+\mathbf{B}_1\cdot\mathbf{\nabla}_\perp\nabla_\perp^2\psi_1=0,
\end{equation}
\begin{equation}\label{eq:psi2}
\partial_t\psi_1=B_c\partial_y\phi_2,
\end{equation}
where (\ref{eq:exden}) to the same order gives
\begin{equation}\label{eq:den2}
\rho_2=\frac{\rho_0'}{B_c}\psi_2+\frac{1}{2}\frac{\rho_0''}{B_c^2}\psi_1^2.
\end{equation}
Using $\psi_1$ from (\ref{eq:psi1sol}), $\phi_2$ can be solved for in (\ref{eq:psi2}) to obtain
\begin{equation}\label{eq:phi2sol}
\phi_2(x,y,t)=\frac{1}{kB_c}\frac{dA(t)}{dt}\zeta_1(x)\sin(ky)+\bar{\phi}_2(x,t),
\end{equation}
where $\bar{\phi}_2(x,t)$ is a constant of integration.

Substituting (\ref{eq:den2}) into (\ref{eq:phi2}) results in an equation for $\psi_2$,
\begin{equation}\label{eq:psi2eig}
\mathcal{L}(\psi_2)=-\frac{g}{B_c^3}\rho_0''\partial_y(\psi_1^2),
\end{equation}
where we have used (\ref{eq:phi12}) to simplify the Laplacian. This has a solution of the form
\begin{equation}\label{eq:psi2sol}
\psi_2(x,y,t)=A(t)^2\zeta_2(x)\cos(2ky)+\bar{\psi}_2(x,t),
\end{equation}
where $\bar{\psi}_2(x,t)$ is the homogeneous solution to (\ref{eq:psi2eig}). Substituting (\ref{eq:psi2sol}) into (\ref{eq:psi2eig}), we find $\zeta_2(x)$ satisfies
\begin{equation}\label{eq:zeta2}
\zeta_2''(x)-4k^2\zeta_2(x)+\frac{g}{B_c^2}\rho_0'\zeta_2(x)=-\frac{1}{2}\frac{g}{B_c^3}\rho_0''\zeta_1(x)^2.
\end{equation}

To fully analyze the stability of our system we still have to resolve the time-evolution of $\psi_1$. It is also important to solve for $\bar{\psi}_2$ and $\bar{\phi}_2$ to make sure that those terms are well-behaved.

\subsection{\label{sec:third}Third order equations}
As was done previously in lower orders, we match terms of $\mathcal{O}(\epsilon^3)$ in (\ref{eq:phi}) and (\ref{eq:psi}). The resulting higher order equations are 
\begin{align}\label{eq:phi3}
\partial_t&(\rho_0\nabla_\perp^2\phi_2+\rho_0'\phi_2')=B_c\partial_y\nabla_\perp^2\psi_3+b_2\partial_y\nabla_\perp^2\psi_1\notag\\&{} + g\partial_y\rho_3+\mathbf{B}_1\cdot\mathbf{\nabla}_\perp\nabla_\perp^2\psi_2+ \mathbf{B}_2\cdot\mathbf{\nabla}_\perp\nabla_\perp^2\psi_1
\end{align}
\begin{equation}\label{eq:psi3}
\partial_t\psi_2=B_c\partial_y\phi_3+\mathbf{B}_1\cdot\mathbf{\nabla}_\perp\phi_2
\end{equation}
along with
\begin{equation}\label{eq:den3}
\rho_3=\frac{\rho_0'}{B_c}\psi_3-\frac{\rho_0'b_2}{B_c^2}\psi_1+\frac{\rho''_0}{B_c^2}\psi_1\psi_2+\frac{1}{6}\frac{\rho_0'''}{B_c^3}\psi_1^3
\end{equation}
from (\ref{eq:exden}).

Integrating (\ref{eq:phi3}) over one period in $y$ we find that $\bar{\phi}_2$ is not driven by $\psi_1$ so we can set
\begin{equation}\label{eq:phibarsol}
\bar{\phi}_2(x,t)=0
\end{equation}
without loss of generality. No zonal flows are generated in the system when creating a periodic perturbation in the density. However, averaging (\ref{eq:psi3}) over $y$ we find that zonal fields are generated according to
\begin{equation}\label{eq:psibarsol}
\bar{\psi}_2(x,t)=\frac{1}{2}\frac{1}{B_c}A(t)^2\zeta_1(x)\zeta_1'(x).
\end{equation}
For a given $k$ and $\rho_0$ the system is now solved up to second order with the exception of the time-evolution $A(t)$. The variables $\psi_1$, $\phi_1$, $\psi_2$, and $\phi_2$ are defined by (\ref{eq:psi1sol}), (\ref{eq:phi1sol}), (\ref{eq:psi2sol}), and (\ref{eq:phi2sol}), respectively. We can solve for $\zeta_1$ using (\ref{eq:eig}) and then for $\zeta_2$ using (\ref{eq:zeta2}). The $y$-independent terms $\bar{\phi}_2$ and $\bar{\psi}_2$ are given by (\ref{eq:phibarsol}) and (\ref{eq:psibarsol}).

To solve for $A(t)$ we need to simplify (\ref{eq:phi3}) by making use of (\ref{eq:phi12}), (\ref{eq:psi2}), and (\ref{eq:den3}). After some algebra (\ref{eq:phi3}) takes the form
\begin{align}\label{eq:phi3psi}
\frac{1}{k^2B_c}&\partial_t^2(\frac{g}{B_c^2}\rho_0\rho_0'\partial_y\psi_1-\rho_0'\partial_y\psi_1')=\notag\\&B_c\mathcal{L}(\psi_3)-2\frac{g}{B_c^2}b_2\rho_0'\partial_y\psi_1+\mathcal{F}[\psi_1,\psi_2],
\end{align}
where exact details of the functional $\mathcal{F}$ is suppressed here for clarity but is shown in Appendix A. We can extract a time-evolution equation by substituting (\ref{eq:psi1sol}), (\ref{eq:psi2sol}), and (\ref{eq:psibarsol}) into the above equation and applying the operator $\int\!\!dx\,\zeta_1(x)\!\int\!\!d(cos(ky))$ evaluated over all space. This operation will annihilate the $\psi_3$ term and any higher order harmonics.

After simplification (see Appendix A), we arrive at the equation for $A(t)$
\begin{align}\label{eq:dimtime}
\frac{1}{k^2B_c}\langle \rho_0\rho_0'\zeta_1^2\rangle&\frac{d^2}{dt^2}A(t)=-2b_2\langle \rho_0'\zeta_1^2\rangle A(t)\notag\\
&+\left(\langle \rho_0''\zeta_1^2\zeta_2\rangle-\frac{1}{4}\frac{1}{B_c}\langle \rho_0'\zeta_1^2(\zeta_1^2)''\rangle\right.\notag\\
&\left.\quad-\frac{1}{4}\frac{B_c}{g}\langle\zeta_1^2(\zeta_1^2)''''\rangle\right)A(t)^3,
\end{align}
where the angled brackets are defined as
\[
\langle f\rangle\equiv \frac{1}{L_\rho}\int\!\!dx\,f(x)
\]
with $L_\rho^{-1}\equiv \rho_0'/\rho_0$ evaluated at $x=0$. We can simplify this further by letting
\begin{align*}
x&\rightarrow \chi L_\rho, \\ \rho_0(x)&\rightarrow \rho_0(0)\rho(\chi), \\ \zeta_1(x)&\rightarrow Z_1(\chi),\\
\zeta_2(x)&\rightarrow Z_2(\chi)/(B_cL_\rho),
\end{align*}
in order to introduce dimensionless versions of the variables $x$ and $\rho_0$, and have $A$ with dimensions of $\psi$. Applying this normalization to (\ref{eq:dimtime}) we get
\begin{equation}\label{eq:time}
\frac{1}{k^2V_{Ac}^2}\frac{d^2}{dt^2}A(t)=-2\frac{b_2}{B_c}c_1A(t)+\frac{c_3}{B_c^2L_\rho^2}A(t)^3
\end{equation}
where $V_{Ac}^2\equiv B_c^2/\rho_0(0)$ and
\begin{equation}\label{eq:coeff1}
c_1=\frac{\langle \rho'Z_1^2 \rangle}{\langle \rho\rho'Z_1^2\rangle},
\end{equation}
\begin{align}\label{eq:coeff2}
c_3=\frac{\langle \rho''Z_1^2Z_2\rangle}{\langle \rho\rho'Z_1^2\rangle}-\frac{1}{4}&\frac{\langle \rho'Z_1^2(Z_1^2)''\rangle}{\langle \rho\rho'Z_1^2\rangle}\notag\\ {}&-\frac{1}{4}\frac{V_{Ac}^2}{gL_\rho}\frac{\langle Z_1^2(Z_1^2)''''\rangle}{\langle \rho\rho'Z_1^2\rangle},
\end{align}
where the primes and brackets now denote derivatives and integrals in $\chi$. Using the same normalization on (\ref{eq:eig}) and (\ref{eq:zeta2}) we get the following equations,
\begin{equation}\label{eq:zeta1nodim}
Z_1''-k^2L_\rho^2Z_1+\frac{gL_\rho}{V_{Ac}^2}\rho'Z_1=0,
\end{equation}
\begin{equation}\label{eq:zeta2nodim}
Z_2''-4k^2L_\rho^2Z_2+\frac{gL_\rho}{V_{Ac}^2}\rho'Z_2=-\frac{1}{2}\frac{gL_\rho}{V_{Ac}^2}\rho''Z_1^2,
\end{equation}
for the dimension-free $Z_1$ and $Z_2$.

The time-evolution equation (\ref{eq:time}) closes the system and we can fully determine the first and second order perturbations, $\psi_1$ and $\psi_2$ defined by (\ref{eq:psi1sol}) and (\ref{eq:psi2sol}), for a given $k$, $\rho_0$, and $b_2$. This is achieved by first solving the eigenvalue problem (\ref{eq:zeta1nodim}) and using the solution for $Z_1$ and $B_c$ to solve for $Z_2$ using (\ref{eq:zeta2nodim}), and finally determining the coefficients (\ref{eq:coeff1}), (\ref{eq:coeff2}) and solving for $A(t)$ in (\ref{eq:time}).

The coefficient $c_1$ is a positive number for $\rho'>0$, and so the linear stability of the system is determined by the sign of $b_2$. This result agrees with the linear theory. However, the overall nonlinear stability of the system is going to be determined largely from the sign of $c_3$ compared to the sign of $b_2$.

\subsection{\label{sec:largek}Short wavelength limit}
We can analytically solve (\ref{eq:zeta1nodim}) for the case $kL_\rho\gg 1$ in which regime the cells are elongated in $x$-direction but still shorter than the scale of the gradient, i.e.,
\begin{equation*}
kL_\rho \gg \chi^{-1} \gg 1.
\end{equation*}
With this scaling we can approximate $\rho'(\chi)$ to be
\begin{equation}
\rho'(\chi)\approx 1-\frac{\chi^2}{2}.
\end{equation}
Assuming that $gL_\rho/V_{Ac}^2 \sim k^2L_\rho^2$, then from scaling arguments we find that (\ref{eq:zeta1nodim}) has the familiar form of a quantum harmonic oscillator. This has the well-known solution
\begin{equation}\label{eq:z1large}
Z_1(\chi)=\hat{Z}_1\exp\left(-\frac{kL_\rho}{2\sqrt{2}}\chi^2\right),
\end{equation}
\begin{equation}\label{eq:z1eiglarge}
k^2L_\rho^2 = \frac{gL_\rho}{V_{Ac}^2}\left(1-\frac{1}{\sqrt{2}}\frac{1}{kL_\rho}\right),
\end{equation}
for the ground state. This solution is correct only for $kL_\rho\gg1$ and the solution for the ``energy'' adds a small correction to the initial assumption. Using the same scaling, to lowest order, (\ref{eq:zeta2nodim}) has the solution
\begin{equation}\label{eq:z2large}
Z_2(\chi)=-\frac{1}{6}\chi Z_1(\chi)^2.
\end{equation}

The time-evolution equation (\ref{eq:time}) can be simplified in the $kL_\rho\gg 1$ limit by substituting the solutions (\ref{eq:z1large})-(\ref{eq:z2large}) in the coefficients (\ref{eq:coeff1}) and (\ref{eq:coeff2}). After simplification we arrive at the following values for the coefficients
\begin{equation}\label{eq:coeffslarge}
c_1=1, \quad c_3=\frac{1}{8}kL_\rho
\end{equation}
where we only kept the largest terms and have assumed that $\hat{Z}_1=1$.

The above result implies that even if $b_2>0$, if the initial amplitude $A_0\equiv A(0)$ is such that
\begin{equation}\label{eq:A0large}
\frac{A_0}{B_cL_\rho} > 4\sqrt{\frac{b_2}{B_c}\frac{1}{kL_\rho}},
\end{equation}
then the system will be nonlinearly unstable and the amplitude will increase without bound. Furthermore, for $b_2<0$ the instability grows faster than predicted from linear theory and any small perturbation will continue to grow larger without saturation.

With the solution for the eigenmode we can check the ratio between the spatial scale of the perturbation, characterized by the displacement in the $x$-direction $\xi_x$, and the width of the eigenmode
\begin{equation}\label{eq:Delta}
\Delta\sim\sqrt{\frac{L_\rho}{k}},
\end{equation}
given by (\ref{eq:z1large}). The displacement is related to the velocity such that $\partial_t\xi_x\sim u_x$, and from (\ref{eq:psi2}) we get that $\partial_tA\sim B_c u_{x2}$, which implies that $A\sim B_c\xi_x$. Substituting for $A$ using (\ref{eq:A0large}) gives us a scale for the displacement,
\begin{equation}
\xi_x\sim\sqrt{\frac{b_2}{B_c}\frac{L_\rho}{k}}
\end{equation}
which yields
\begin{equation}\label{eq:scaleratio}
\frac{\xi_x}{\Delta}\sim\sqrt{\frac{b_2}{B_c}}
\end{equation}
for the ratio of the two scale lengths. As should be expected, the spatial size of the amplitude required to be nonlinearly unstable is much smaller than the width of the eigenmode.

\begin{figure}
\includegraphics[width=0.48\textwidth]{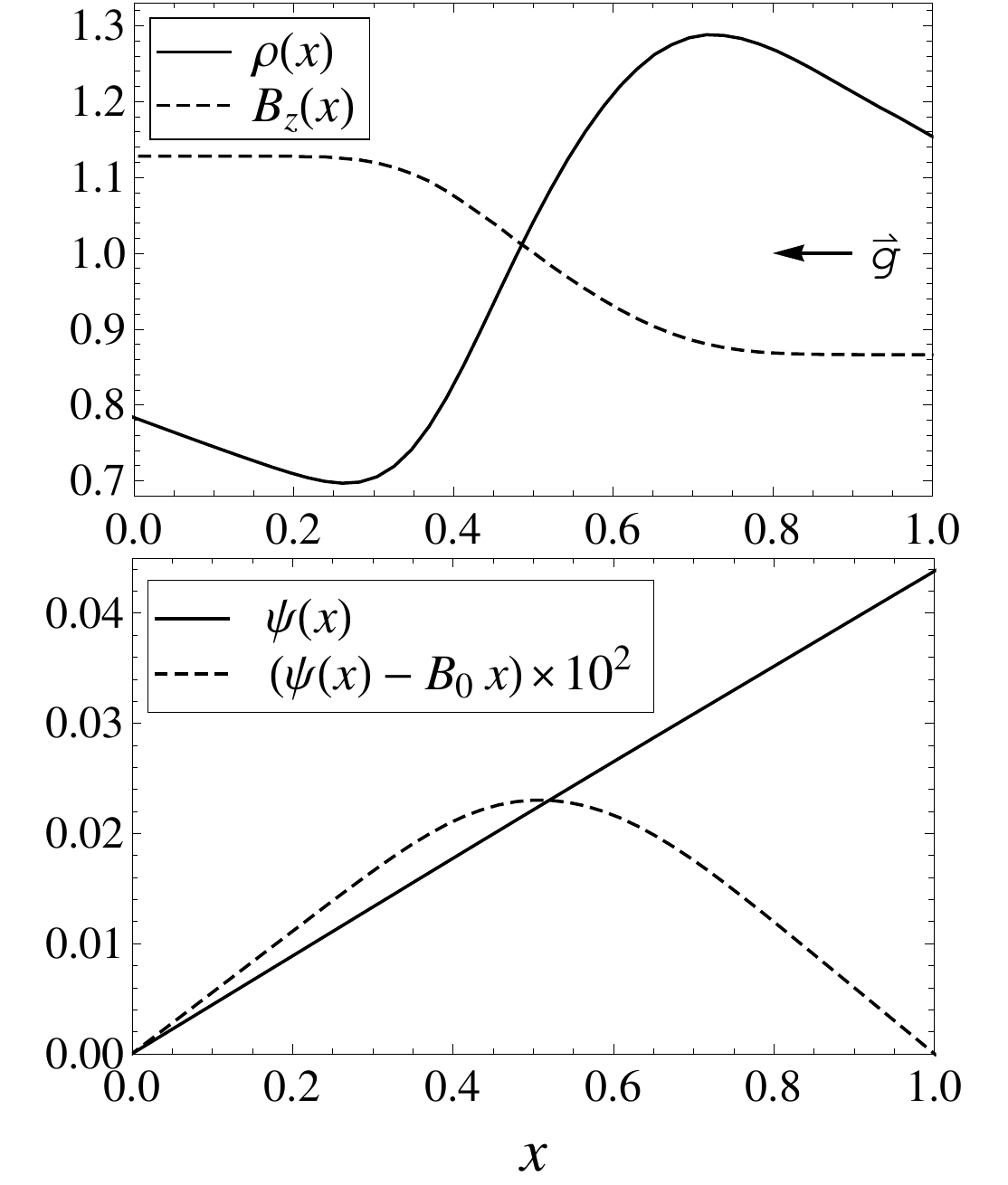}
\caption{\label{fig:profile} The equilibrium profiles for the background field $B_z$, the density $\rho$, and the magnetic streamfunction $\psi$ along with the difference from constant field.}
\end{figure}

\section{\label{sec:num}Numerical Simulation}
To confirm this result, we used a two-dimensional code that solves the fully compressional equations (see Appendix B). The variables $\rho$, $\rho\mathbf{u}$, $\psi$, and $B_z$ are solved numerically and stepped in time. We set $B_z \gg |B_\perp|$ so the equations are effectively reduced. The code is dissipative so we introduced source terms in the density in order to maintain a steady state profile suitable for our model. The sourcing, although weak, results in a profile for $B_y(x)$. To compare with analytic theory, we wish to keep $B_y$ approximately constant. Thus, we allowed $B_z$ to resistively relax at a somewhat slower rate than $B_y$ in the equilibrium.

The system is normalized so that initially $V_{Az}=1$ and $L_x=1$, where $L_x$ is the height of the box. We used hardwall, free-slip boundary conditions for the top and bottom walls and periodic boundary conditions for the sides. The periodic boundary conditions discretize the system so that the only wavenumbers allowed are integer multiples of $2\pi/L_y$, where $L_y$ is the width of the box. From (\ref{eq:disp}) we know that the lower modes are the most unstable, so to study the case with $kL_\rho\gg 1$, i.e. short wavelength, we selected $L_y$ such that the minimum value for $kL_\rho$ satisfies this condition. By choosing $k=2\pi/L_y$ we can satisfy the marginality condition by adjusting $B_0$ and/or $g$ such that $kV_{Ay}\approx\gamma_g$ for the minimum mode. We set $L_y=0.5$, and from the density profile we have $L_\rho=\rho_0/\rho_0'\approx0.4$ and so we satisfy the condition
\begin{equation*}
kL_\rho\approx5.03 \gg 1
\end{equation*}
which is necessary to compare with the analytical result from Section \ref{sec:largek}. We attempted to run tests with a larger value of $k$ by decreasing $L_y$, but the code was numerically unstable for smaller box widths.

To generate the equilibrium we initialize $\psi$ to $B_0x$ and let the system reach an equilibrium which is steady state. The density source term results in a weak flow in the $x$-direction. This flow scales with the diffusion, so a minimal, numerically-stable value for the diffusion is chosen to minimize its effect. The equilibrium profiles for the density and the background field generated are shown in Fig.~\ref{fig:profile}. It is important to note that the equilibrium profile for the density does not have $\rho_0'\rightarrow 0$ at the boundaries. The boundary conditions imply that $\rho_0'\rightarrow -g\rho_0$ at the wall.

\begin{figure}
\includegraphics[width=0.48\textwidth]{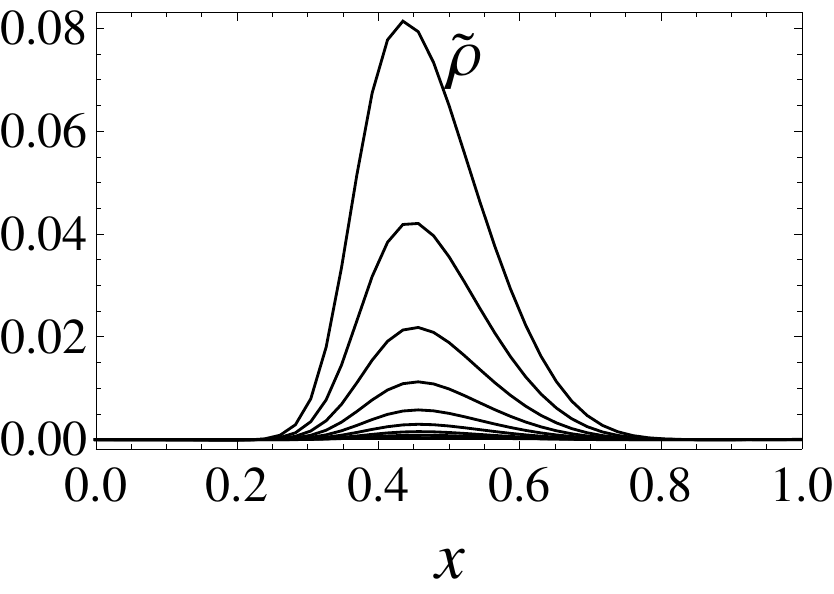}
\caption{\label{fig:deneigx} The linear growth of an unstable localized mode cut at $y\approx 0.26$ and for $t\leq 60\tau_A$. Time traces separated by $t\approx6\tau_A$ are shown.}
\end{figure}

After the equilibrium is made, a density perturbation is introduced with $\tilde{\rho}(x,y)=a_0\cos(ky)$. From (\ref{eq:den1}) we can relate the density perturbation amplitude, $a(t)$, to the perturbation amplitude of $\psi$, i.e. $a=\rho_0'A/B_c$. In Fig.~\ref{fig:deneigx} we show the resulting unstable eigenmode developing for the density. For tests done with $B_0$ far away from marginality, i.e. $|b_2/B_c|\approx50\%$, there was excellent agreement for the growth rate/frequency in the simulation with (\ref{eq:disp}). The theory predicts that there will be nonlinear coupling to the mode with wavenumber $2k$, so it is important that this mode and higher modes are allowed. Since the diffusivity is weak, it is ensured that this is the case.

Since we can adjust both $B_0$ and $g$ to achieve marginal stability, we decided to fix the value of $g$ at $0.15$, and adjust $B_0$. With this value of $g$ we expect that $B_c \approx 0.05$ based on (\ref{eq:z1eiglarge}). However, we found that an equilibrium with $B_0=0.05$ is stable to perturbations as large as $a_0=10^{-1}$ in the simulation. We decreased the strength of the transverse field until it became unstable to perturbations with $a_0=10^{-4}$. This value was at $B_0\approx0.0438$ and we took this to be the critical value of the transverse field for the numerical simulation. Since the critical amplitude scales like the square root of the deviation from marginality, we are limited to perturbations only as small as $10^{-4}$ otherwise smaller perturbations would have meant having deviations that are close to the limits of our computational power. 

\begin{figure}
\includegraphics[width=0.48\textwidth]{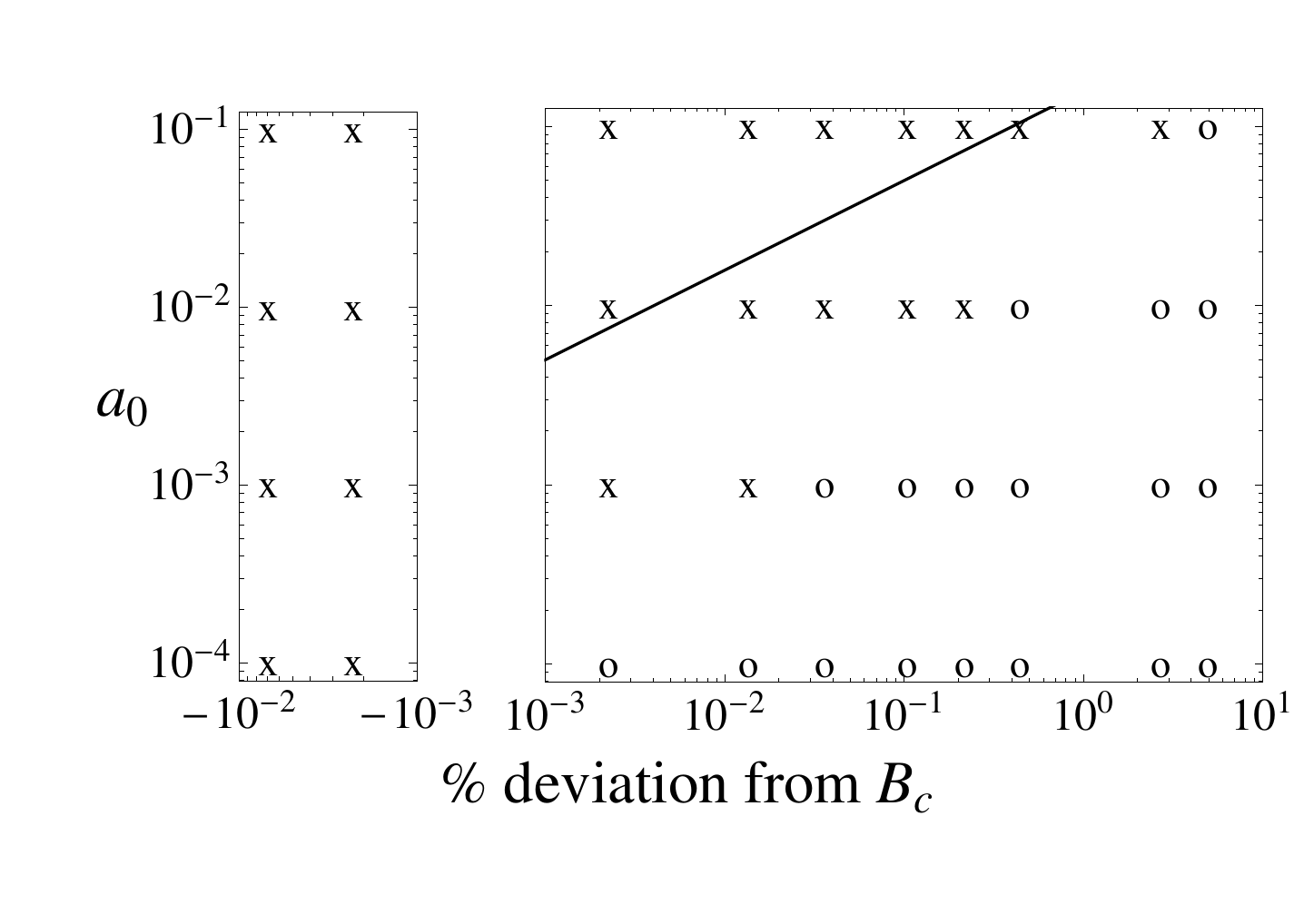}
\caption{\label{fig:result} Result of stability test for a range of deviations from $B_c$ and magnitude of perturbation, $a_0$. Stable and unstable results are denoted by a circle or a cross, respectively. The solid line is the theoretical boundary.}
\end{figure}

We created multiple equilibria with different transverse field strength within 10\% of the numerical critical field strength. These equilibria were then perturbed with $a_0$ of different orders of magnitude. The result of the test is shown in Fig.~\ref{fig:result} where circles and crosses mark stable and unstable points, respectively, and the solid line is for $a_0 = 4\rho_0'\sqrt{(b_2/B_c)(L_\rho/k)}$, from our theory, using the parameters from the numerical simulation. The slope of the theory line seems consistent with the numerical data, however, the theory requires larger $a_0$ for nonlinear instability. This inconsistency could be due to the diffusion in the code and, in particular, the resistivity may allow for slippage in the magnetic field lines which can shift the stability boundary at marginal stability. We can calculate the scale size of this shift based on the values used in the simulation (see Appendix B),
\begin{equation}
\frac{\eta/\Delta^2}{kV_{Ay}}\approx 2.5\%.
\end{equation}
This implies that there could be a shift in $B_c$ of order $\sqrt{b_2/B_c}$. At marginal stability, even small diffusion can cause significant shifts in stable-unstable boundaries. However, this implies a shift in $B_c$; it is harder to explain why resistivity results in a nonlinear instability at large amplitude of perturbation. It is possible that diffusive effects may affect the critical amplitude for nonlinear instability, but the existence of a nonlinear instability phenomenon is harder to explain as a diffusive effect.

\begin{figure}
\includegraphics[width=0.48\textwidth]{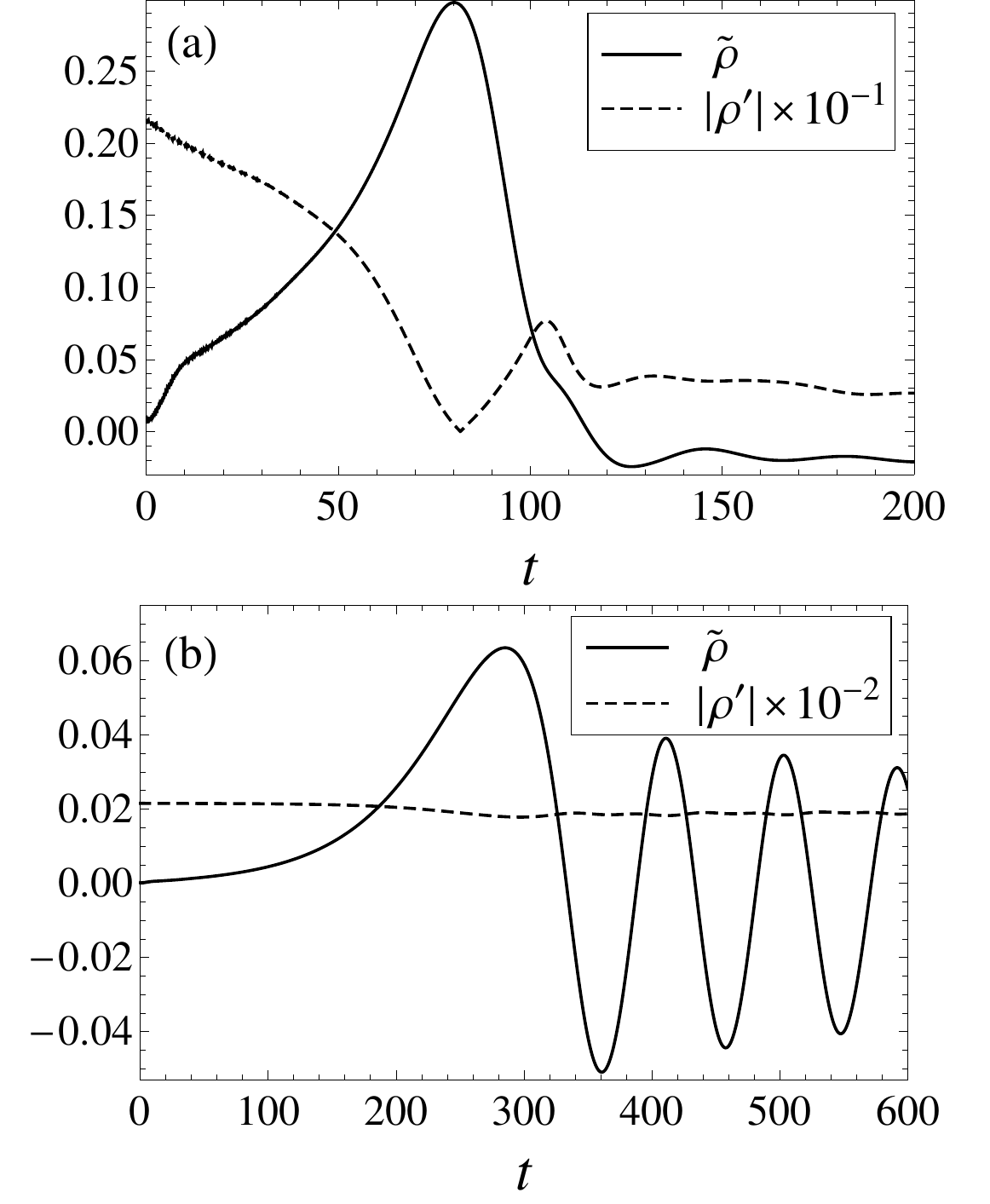}
\caption{\label{fig:traces} Time trace of the amplitude of density perturbations $\tilde{\rho}$ (solid line) and $x$ derivative of the density $\rho'$ (dashed line) for $b_2/B_c\approx0.04\%$ with (a) $a_0=10^{-2}$ and (b) $a_0=10^{-4}$.} 
\end{figure}

\begin{figure}
\includegraphics[width=0.48\textwidth]{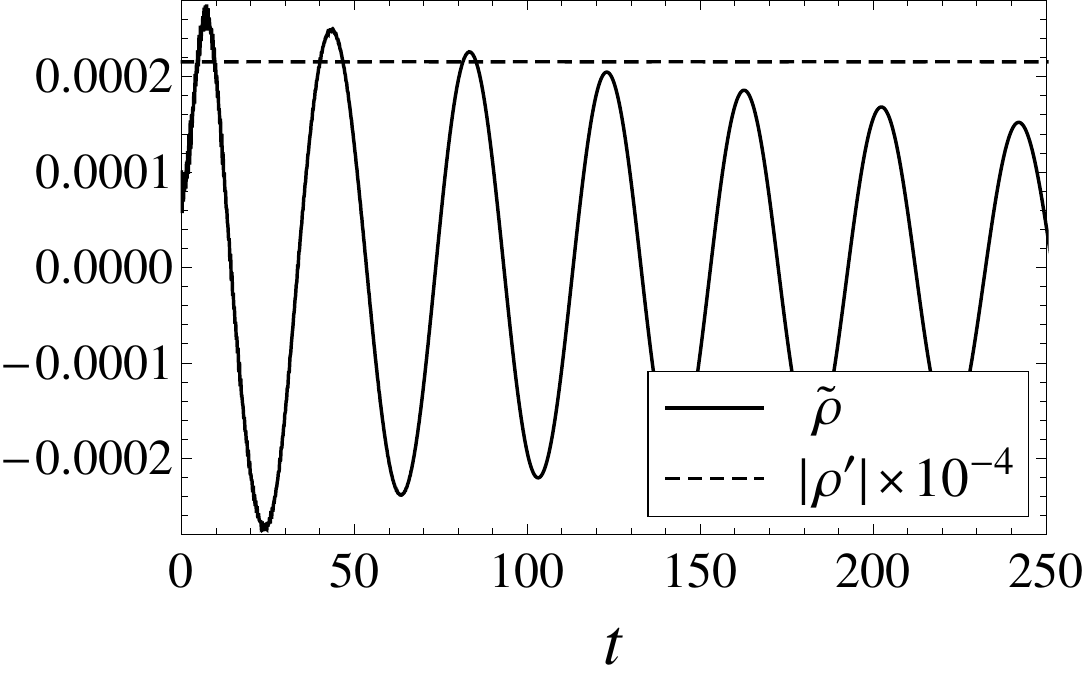}
\caption{\label{fig:Sstrace} Time trace of the amplitude of density perturbations $\tilde{\rho}$ (solid line) and $x$ derivative of the density $\rho'$ (dashed line) for $b_2/B_c\approx10\%$ with $a_0=10^{-4}$.} 
\end{figure}

In addition to checking the perturbations for a growing linear mode, we also check the time trace of the amplitude for nonlinear effects. In Fig.~\ref{fig:traces} we show a time trace of the amplitude of $\tilde{\rho}$, $a(t)$, for the same $B_0$ but different $a_0$. We can see that the behaviours are different for the two cases. In the unstable case, Fig.~\ref{fig:traces}a, the density perturbations become very large quickly and eventually dissipate after it hits the boundaries ($t\lesssim100\tau_A$). The time trace of $\rho'$ shows that the density profile flattens out ($\rho'\rightarrow0$) after reaching a peak. So, even though our analysis in Sec.~\ref{sec:theory} is only valid as long as $A\lesssim\epsilon$ we can see from the trace that it continues beyond this limit until the profile collapses. The stable case, Fig.~\ref{fig:traces}b, has an initial growth eventually hitting a peak and then has stable oscillations. Even though the amplitude increases some, it is still small and the density profile holds. This can be seen from the fact that $\rho'$ is staying constant the entire time. We can see in Fig.~\ref{fig:Sstrace} that as we increase $b_2/B_c$ further from marginality, this initial growth decreases in magnitude. It also develops faster and has more noise that is indicative of a transient oscillatory mode.

\section{\label{sec:summary}Summary and Conclusions}
In this paper, we studied the nonlinear behaviour of a marginally stable interchange system. We used the reduced equations to find an analytic solution near marginality given a density profile, $\rho_0(x)$, deviation from marginality, $b_2$, and wavenumber of perturbation, $k$, of the B-field. The result is a nonlinear differential equation for the amplitude of the density perturbations as a function of time. The threshold for nonlinear instability is dependent on the above quantities, along with $g$. The principal finding of this paper is that marginally stable interchange modes in a magnetized plasma can be nonlinearly unstable for large enough initial perturbations. We arrived at this result from a systematic asymptotic expansion about marginality in the smallness parameter, $|b_2/B_c|^{1/2}$, carried out to third order. The first order solution can be found using the linear eigenvalue problem. This solution is then used as a source for the second order problem. The third order analysis yields the equation for the time-dependence of the perturbation. We found that the stability of the solution can be determined by calculating the coefficient of the nonlinear term in the differential equation. This is a nontrivial task for a general perturbation, but we could analytically solve this in the short wavelength limit. In this limit we found that the nonlinear coefficient had a positive sign. This meant that in the linearly stable case ($b_2>0$) it was possible to be nonlinearly unstable if the initial perturbation was large enough. We found the critical amplitude to be proportional to $\sqrt{b_2}$.
 
A nonlinear numerical MHD simulation fully confirms the analytic result. We have used a numerical simulation of the nonlinear, full, compressible, MHD equations with small dissipation to verify our analytical result. We showed very good agreement between the simulation and the theory for deviations, $b_2$, from $B_c$ of up to 10\%. The numerical results show that in the short wavelength limit the system is nonlinearly unstable. There is some disagreement in the time evolution of the density with the analytical result, but this is possible since the analytic calculation is for an ideal system with no dissipation. We also discussed why a shift in $B_c$ for the linear instability threshold, due to dissipation, is possible at marginal stability and how it is harder to explain why the nonlinear result has an amplitude dependent stability. Furthermore, the dependence is cubic so the mode grows without bound once it is unstable. This is even harder to explain as a resistive effect.

It should be noted that the fully analytic calculation is facilitated by using a very simple form (a constant) for the transverse stabilizing magnetic field. So, while the conclusions of this paper seem to be on solid ground, the application of these findings to various systems, to the extent that the transverse $B$ field of this paper is very special, must be appropriately qualified. For example, in tokamaks and stellarators, the interchange mode arises on rational surfaces which corresponds to a slab model with a sheared magnetic field vanishing at $x=0$. In the solar coronal case, line-tying is an important characteristic absent in our simple case. Nonetheless, the conclusions are sufficiently dissimilar as to indicate further investigation. Thus, for example, a neighboring nonlinear saturated state for the interchange mode was found in Refs.~\onlinecite{Waelbroeck,Beklemishev} -- whereas the corresponding result in our case, for $b_2<0$, indicates a robustly growing mode with no nonlinear saturation. Of course, the transverse magnetic field in these papers was a sheared field with a rational surface for the unstable wave mode. Attempting a marginal stability analysis for sheared field, similar to that used in the present paper, is not straightforward. The fact that the sheared field goes to zero as $x$ goes to zero means that a new inner ordering is required, which makes the calculation more involved.

Our results are more consistent with the nonlinear instability found in Ref.~\onlinecite{Cowley} where the authors were also in the parameter range with $k_\perp \gg 1$, $\Delta_x \sim k_\perp^{-1/2}$, and $\xi_x \ll \Delta_x$. It should be noted that their analysis was for the 3D line-tied $g$ mode with no transverse field at marginal stability. Even so, the suprising result is that in both cases the system takes off once it becomes nonlinearly unstable. This occurs even when the linear term is stabilizing. The primary difference between the results is the amplitude dependence of the nonlinear term. In Ref.~\onlinecite{Cowley} the nonlinear term has a quadratic dependence, while our analysis yields a cubic dependence on amplitude. If we construct an effective potential, we observe that the result from Ref.~\onlinecite{Cowley} indicates a dependence on the sign of the perturbation at the metastable boundary, while our potential is symmetric in $A$. Another difference is that the result in Ref.~\onlinecite{Cowley} was somewhat mitigated by Refs.~\onlinecite{Zhu1,Zhu2} in that the latter papers argued that the ordering giving nonlinear growth would break down at small amplitudes before the instability fully takes off. In our case, our numerical simulations seem to show, in agreement with analytic constraints, that the nonlinear instability growth continues without bound and the theory only fails when $A\sim\mathcal{O}(1)$ (as saturation is reached).

Our results could also be relevant to tokamak ballooning modes to the extent that these modes are stabilized by an ``average minimum-$B$ well'' and thus always have some parallel wavenumber. Work is in progress to quantify this better. Finally, our results also indicate a closer look at interchange stability in stellarators, presumably in average minimum-$B$ stabilized systems.

Further investigation is necessary to answer some questions regarding the results found in this paper. The transient initial growth in the time traces, mentioned in Sec.~\ref{sec:num}, needs to be explained. The change in the growth rate as the system gets closer to marginal stability, with $b_2<0$, needs to be investigated and compared to the results from Ref.~\onlinecite{Gupta}.

\section*{Acknowledgments}
This work was supported by the U.S. Department of Energy. JB and ABH dedicate this manuscript to the memory of Dr. Parvez Guzdar.

\renewcommand{\theequation}{A\arabic{equation}}
\setcounter{equation}{0}
\appendix*
\section*{APPENDIX A}
\subsection*{Derivation of (\ref{eq:time})-(\ref{eq:coeff2})}

In simplifying (\ref{eq:phi3psi}), we found the functional, $\mathcal{F}[\psi_1,\psi_2]$, to be
\begin{align}\label{eq:func}
\mathcal{F}[\psi_1,\psi_2]=&\frac{g}{B_c^2}\rho_0''\partial_y(\psi_1\psi_2)+\frac{1}{6}\frac{g}{B_c^3}\rho_0'''\partial_y(\psi_1^3)\notag\\&+\mathbf{B}_1\cdot\mathbf{\nabla}_\perp\nabla_\perp^2\psi_2+ \mathbf{B}_2\cdot\mathbf{\nabla}_\perp\nabla_\perp^2\psi_1.
\end{align}
The above equation can be simplified by writing $\psi_1$ and $\psi_2$ a certain way. From (\ref{eq:psi2sol}) we can write 
\begin{equation}\label{eq:apppsi2}
\psi_2=\tilde{\psi}_2+\bar{\psi}_2
\end{equation}
where
\begin{equation}\label{eq:psi2twid}
\tilde{\psi}_2=A(t)^2\zeta_2(x)\cos(2ky).
\end{equation}
Writing $\psi_2$ in this way, we get the following results
\begin{equation}\label{eq:app0}
\partial_y\psi_2=\partial_y\tilde{\psi}_2,
\end{equation}
\begin{equation}\label{eq:app1}
\nabla_\perp^2\tilde{\psi}_2=-\frac{g}{B_c^2}\rho_0'\tilde{\psi}_2-\frac{g}{B_c^3}\rho_0''\widetilde{\psi_1^2},
\end{equation}
where we have written $\psi_1^2=\widetilde{\psi_1^2}+\overline{\psi_1^2}$ and
\begin{equation}\label{eq:psi1widetilde}
\widetilde{\psi_1^2}=\frac{1}{2}A(t)^2\zeta_1(x)^2\cos(2ky),
\end{equation}
\begin{equation}\label{eq:psi1overline}
\overline{\psi_1^2}=\frac{1}{2}A(t)^2\zeta_1(x)^2.
\end{equation}
The result (\ref{eq:app1}) can be derived by multiplying (\ref{eq:zeta2}) with $A(t)^2\cos(2ky)$ and recombining the terms. Similarly, if we multiply (\ref{eq:eig}) by $A(t)\cos(ky)$, we find that
\begin{align}\label{eq:app2}
\nabla_\perp^2\psi_1=-\frac{g}{B_c^2}\rho_0'\psi_1.
\end{align}

Since $\mathbf{B}\cdot\mathbf{\nabla}_\perp\psi=0$ for all orders, we get that
\begin{align}
\mathbf{B_1}\cdot\mathbf{\nabla}_\perp\psi_1^2&=2\psi_1(\mathbf{B_1}\cdot\mathbf{\nabla}_\perp\psi_1)\notag\\
&=0\notag\\
&=\mathbf{B_1}\cdot\mathbf{\nabla}_\perp(\widetilde{\psi_1^2}+\overline{\psi_1^2}),
\end{align}
and therefore
\begin{align}\label{eq:app3}
\mathbf{B_1}\cdot\mathbf{\nabla}_\perp\widetilde{\psi_1^2}&=-\mathbf{B_1}\cdot\mathbf{\nabla}_\perp\overline{\psi_1^2}\notag\\& = \partial_y\psi_1\overline{\psi_1^2}~',
\end{align}
Similarly, since
\begin{align}
\mathbf{B_1}\cdot\mathbf{\nabla}_\perp\psi_2+\mathbf{B_2}\cdot\mathbf{\nabla}_\perp\psi_1&=0\notag\\
&=\mathbf{B_1}\cdot\mathbf{\nabla}_\perp(\tilde{\psi}_2+\bar{\psi}_2)\notag\\& \quad\quad+\mathbf{B_2}\cdot\mathbf{\nabla}_\perp\psi_1
\end{align}
then it follows that
\begin{align}\label{eq:app4}
\mathbf{B_1}\cdot\mathbf{\nabla}_\perp\tilde{\psi}_2+\mathbf{B_2}\cdot\mathbf{\nabla}_\perp\psi_1&= -\mathbf{B_1}\cdot\mathbf{\nabla}_\perp\bar{\psi}_2\notag\\&=\partial_y\psi_1\bar{\psi}_2'.
\end{align}

We can now simplify the last two terms in (\ref{eq:func}). Using (\ref{eq:apppsi2}) we have
\begin{align}\label{eq:B1psi2}
\mathbf{B}_1\cdot\mathbf{\nabla}_\perp\nabla_\perp^2\psi_2&=\mathbf{B}_1\cdot\mathbf{\nabla}_\perp\nabla_\perp^2(\tilde{\psi}_2 +\bar{\psi}_2)\notag\\
&=\mathbf{B}_1\cdot\mathbf{\nabla}_\perp(-\frac{g}{B_c^2}\rho_0'\tilde{\psi}_2-\frac{g}{B_c^3}\rho_0''\widetilde{\psi_1^2})\notag\\&\quad\quad+\mathbf{B}_1\cdot\mathbf{\nabla}_\perp\bar{\psi_2}''\notag\\
&=-\frac{g}{B_c^2}\rho_0'\mathbf{B}_1\cdot\mathbf{\nabla}_\perp\tilde{\psi}_2+\frac{g}{B_c^2}\rho_0''\partial_y\psi_1\tilde{\psi}_2\notag\\
&\quad\quad-\frac{g}{B_c^3}\rho_0''\mathbf{B}_1\cdot\mathbf{\nabla}_\perp\widetilde{\psi_1^2}+\frac{g}{B_c^3}\rho_0'''\partial_y\psi_1\widetilde{\psi_1^2}\notag\\&\quad\quad\quad-\partial_y\psi_1\bar{\psi}_2''',
\end{align}
where we used (\ref{eq:app1}), and took advantage of the fact that $\bar{\psi}_2$ has no $y$ dependence, to remove the Laplacians. Similarly, we use (\ref{eq:app2}) to get
\begin{align}\label{eq:B2psi1}
\mathbf{B}_2\cdot\mathbf{\nabla}_\perp\nabla_\perp^2\psi_1&=\mathbf{B}_2\cdot\mathbf{\nabla}_\perp(-\frac{g}{B_c^2}\rho_0'\psi_1)\notag\\
&=-\frac{g}{B_c^2}\rho_0'\mathbf{B}_2\cdot\mathbf{\nabla}_\perp\psi_1+\frac{g}{B_c^2}\rho_0''\partial_y\tilde{\psi}_2\psi_1,
\end{align}
where we used (\ref{eq:app0}) to get the second term.

Combining (\ref{eq:B1psi2}) and (\ref{eq:B2psi1}) we can use (\ref{eq:app3}) and (\ref{eq:app4}) to further simplify the terms with a gradient operator. So finally we get
\begin{align}
\mathbf{B}_1\cdot\mathbf{\nabla}_\perp\nabla_\perp^2\psi_2+&\mathbf{B}_2\cdot\mathbf{\nabla}_\perp\nabla_\perp^2\psi_1= -\frac{g}{B_c^2}\rho_0'\partial_y\psi_1\bar{\psi}_2'\notag\\
&\quad-\frac{g}{B_c^3}\rho_0''\partial_y\psi_1\overline{\psi_1^2}~'+\frac{g}{B_c^2}\rho_0''\partial_y(\psi_1\tilde{\psi}_2)\notag\\
&\quad\quad+\frac{g}{B_c^3}\rho_0'''\partial_y\psi_1\widetilde{\psi_1^2}-\partial_y\psi_1\bar{\psi}_2'''.
\end{align}
We can also rewrite the first term of (\ref{eq:func}),
\begin{align}
\frac{g}{B_c^2}\rho_0''\partial_y(\psi_1\psi_2)=\frac{g}{B_c^2}\rho_0''\partial_y(\psi_1\tilde{\psi}_2)+\frac{g}{B_c^2}\rho_0''\bar{\psi}_2\partial_y\psi_1.
\end{align}

We can now substitute for $\psi_1$, $\bar{\psi}_2$, $\tilde{\psi}_2$, $\widetilde{\psi_1^2}$ and $\overline{\psi_1^2}$ using (\ref{eq:psi1sol}), (\ref{eq:psibarsol}), (\ref{eq:psi2twid}), (\ref{eq:psi1widetilde}), and (\ref{eq:psi1overline}). As described in Section \ref{sec:third}, we use the operator $\int\!\!dx\,\zeta_1(x)\!\int\!\!d(cos(ky))$ on (\ref{eq:phi3psi}) in order to extract the terms that have a $\sin(ky)$ dependence. The other terms will be irrelevant since the integration will evaluate to zero if the dependence doesn't match. And so we find that
\begin{align}\label{eq:app5}
\int\!\!d(cos(ky))&\mathcal{F}[\psi_1,\psi_2]=\pi kA(t)^3\left\{\frac{g}{B_c^2}\rho_0''\zeta_1\zeta_2\notag\right.\\
&-\frac{1}{4}\frac{g}{B_c^3}\left(\rho_0'\zeta_1(\zeta_1^2)''+\rho_0''\zeta_1(\zeta_1^2)'+\frac{1}{2}\rho_0'''\zeta_1^3\right)\notag\\ &\quad-\left.\frac{1}{4}\frac{1}{B_c}\zeta_1(\zeta_1^2)''''\right\}
\end{align}
Finally, we use the operator $\int\!\!dx\,\zeta_1(x)$ on the above equation to get
\begin{align}\label{eq:collect1}
\int\!\!dx\,\zeta_1(x)&\!\int\!\!d(cos(ky))\mathcal{F}[\psi_1,\psi_2]=\pi kL_\rho A(t)^3\times\notag\\
&\left(\frac{g}{B_c^2}\langle \rho_0''\zeta_1^2\zeta_2\rangle-\frac{1}{4}\frac{g}{B_c^3}\langle \rho_0'\zeta_1^2(\zeta_1^2)''\rangle\right.\notag\\
&\quad\left.-\frac{1}{4}\frac{1}{B_c}\langle\zeta_1^2(\zeta_1^2)''''\rangle\right).
\end{align}
We made use of the fact that $\zeta_1(x)$ decays exponentially at the boundaries to combine the three terms proportional to $g/B_c^3$ in (\ref{eq:app5}) into one term through integration by parts.

To complete the derivation of (\ref{eq:time})-(\ref{eq:coeff2}) we still need to simplify the rest of the terms. It is easy to see that after using the annihilation operator then we get
\begin{align}\label{eq:collect2}
\int\!\!dx\,\zeta_1(x)\!\int\!\!d(cos&(ky))(-2\frac{g}{B_c^2}b_2\rho_0'\partial_y\psi_1)=\notag\\
&-2\pi kL_\rho A(t)\frac{g}{B_c^2}b_2\langle \rho_0'\zeta_1^2\rangle.
\end{align}
Applying the same operator, we find that
\begin{align}
\int\!\!dx\,\zeta_1(x)\!&\int\!\!d(cos(ky))B_c\mathcal{L}(\psi_3)\notag\\
&=-kB_c\int\!\!dy\!\int\!\!dx\,\zeta_1\sin(ky)\big(\nabla_\perp^2\psi_3+\frac{g}{B_c^2}\rho_0'\psi_3\big)\notag\\
&=-kB_c\int\!\!dy\!\int\!\!dx\,\bigg(\zeta_1''\sin(ky)\psi_3\notag\\
&\quad\quad+\zeta_1(-k^2\sin(ky))\psi_3+\zeta_1\sin(ky)\frac{g}{B_c^2}\rho_0'\psi_3\bigg)\notag\\
&=-kB_c\int\!\!dy\!\int\!\!dx\,\sin(ky)\psi_3\times\notag\\
&\quad\quad\quad\big(\zeta''-k^2\zeta_1+\frac{g}{B_c^2}\rho_0'\zeta_1\big),
\end{align}
and therefore, using (\ref{eq:eig}),
\begin{align}\label{eq:collect3}
\int\!\!dx\,\zeta_1(x)\!&\int\!\!d(cos(ky))B_c\mathcal{L}(\psi_3)=0.
\end{align}
We, once again, took advantage of the boundary conditions to perform some integration by parts to arrive at the above result. Lastly, the operator on the left-hand side of (\ref{eq:phi3psi}) gives
\begin{align}\label{eq:collect4}
\int\!\!dx\,\zeta_1(x)\!&\int\!\!d(cos(ky))(\frac{g}{B_c^2}\rho_0\rho_0'\partial_y\psi_1-\rho_0'\partial_y\psi_1')\notag\\
&=\pi kL_\rho A(t)(\frac{g}{B_c^2}\langle \rho_0\rho_0'\zeta_1^2\rangle-\langle \rho_0'\zeta_1\zeta_1'\rangle)\notag\\
&=\pi kL_\rho A(t)\frac{g}{B_c^2}\langle \rho_0\rho_0'\zeta_1^2\rangle,
\end{align}
where the second term was thrown away since it evaluates to zero due to the parity of the equilibrium density.

Collecting the terms (\ref{eq:collect1}), (\ref{eq:collect2}), (\ref{eq:collect3}) and (\ref{eq:collect4}) together, we arrive at the (\ref{eq:dimtime}).

\renewcommand{\theequation}{B\arabic{equation}}
\setcounter{equation}{0}
\appendix*
\section*{APPENDIX B}
\subsection*{Description of numerical simulation}

The two-dimensional numerical simulation solves the following equations:
\begin{equation}\label{eq:numrho}
\partial_t\rho+\mathbf{\nabla}_\perp\cdot(\rho\mathbf{u}_\perp)-D_\rho\nabla_\perp^2\rho=S,
\end{equation}
\begin{equation}\label{eq:numuperp}
\partial_t(\rho\mathbf{u}_\perp)+\mathbf{\nabla}_\perp\cdot(\rho\mathbf{u}_\perp\mathbf{u}_\perp)-\mu\nabla_\perp^2(\rho\mathbf{u}_\perp)=\mathbf{F}_\perp,
\end{equation}
\begin{equation}\label{eq:numuz}
\partial_t(\rho u_z)+\mathbf{\nabla}_\perp\cdot(\rho u_z\mathbf{u}_\perp)-\mu\nabla_\perp^2(\rho u_z)=[B_z,\psi],
\end{equation}
\begin{equation}\label{eq:numBz}
\partial_tB_z+\mathbf{\nabla}_\perp\cdot(B_z\mathbf{u}_\perp)-\eta_\perp\nabla_\perp^2B_z=[\psi,u_z],
\end{equation}
\begin{equation}\label{eq:numpsi}
\partial_t\psi+\mathbf{u}_\perp\cdot\mathbf{\nabla}_\perp\psi-\eta\nabla_\perp^2\psi=0,
\end{equation}
where $[f,h]\equiv\partial_xf\partial_yh-\partial_xh\partial_yf$ and
\begin{equation}\label{eq:numFperp}
\mathbf{F}_\perp=-\mathbf{\nabla}_\perp\left(\frac{T_0}{M}\rho+\frac{B_z^2}{2}\right)-\mathbf{\nabla}_\perp\psi\nabla_\perp^2\psi-\rho g\mathbf{\hat{x}},
\end{equation}
\begin{equation}\label{eq:numsource}
S=\eta_\perp S_0\left(e^{-(x-x_1)^2/2\sigma^2}-e^{-(x-x_2)^2/2\sigma^2}\right).
\end{equation}
The system is initialized with $\rho=1$ and $B_z=1$. We use $T_0/M=0.3$ for the temperature and $g=0.15$ for the gravitational acceleration. The Gaussian function sources have amplitude $S_0=4.5$, width $\sigma^2=6.25\times10^{-4}$ and centered around $x_1=0.7$ and $x_2=0.38$ (where $L_x=1$). The values are chosen by trial and error to create a good $\rho_0'(x)$ profile for the simulation. The relative strength of the dissipation terms are as follows:
\begin{align*}
\mu = \eta &=5\times 10^{-4}, \\
\eta_\perp &=10^{-1}\eta, \\
D_\rho &=10^{-3}\mu.
\end{align*}
The dissipation in the density, $D_\rho$, is for numerical stability and is made orders of magnitude smaller than the viscosity $\mu$. As mentioned in Sec.~\ref{sec:num} the $B_z$ resistivity, $\eta_\perp$, is made smaller than $\eta$ and $\mu$ in order to keep $B_y$ approximately constant. The crossfield particle diffusion is set by $\eta_\perp$. Since the time and space scales are normalized to the Alfv\'{e}n speed, $V_{Az}$, and the box size, $L_x$, the above coefficients imply a viscous magnetic Reynolds number of $\simeq 2\times 10^3$ and a Lundquist number (for magnetic diffusion) of $\simeq 2\times 10^4$.

\providecommand{\noopsort}[1]{}\providecommand{\singleletter}[1]{#1}%

\end{document}